\documentclass[12pt]{iopart}

\usepackage{epsf}

\newcommand{\Sgmm}{\Sigma^0}
\newcommand{\Sgm}{$\Sgmm$\ }
\newcommand{\GSgm}{$\Sigma$\ }
\newcommand{\ASgmm}{\overline{\Sigma}^0}
\newcommand{\ASgm}{$\ASgmm$\ }
\newcommand{\Gm}{$\gamma$}
\newcommand{\Lam}{$\Lambda$\ }
\newcommand{\ALam}{$\overline{\Lambda}$\ }
\newcommand{\Ratio}{$\Sgmm/\Lambda$\ }
\newcommand{\Piz}{$\pi^0$\ }
\newcommand{\ptarm}{$p_T^{AP}$}

\begin{document}

\title{Reconstructing \Sgm decays in STAR}

\author{G. Van Buren\dag\ for the STAR Collaboration\footnote{For
the full author list and acknowledgements, see Appendix "Collaborations" in this volume.}
}
\address{\dag\ Dept. of Physics, Brookhaven National Laboratory, Upton, NY 11973-5000 USA}
\ead{gene@bnl.gov}

\begin{abstract}
Typical comparisons of data from nuclear collisions
to particle production models require a caveat for
(anti)\Lam yields from experimental inability
to separate the contributions of those yields from
\GSgm state decays. Recent analysis in STAR is
leading toward resolving the contribution from
excited \GSgm states~\cite{sevil}, but the bulk contribution
comes from electromagnetic decays of the
(anti)\Sgm.

In the STAR detector, photon conversions into $e^+e^-$
pairs in the detector material have been used to
identify photons from \Piz decays~\cite{pi0}. A similar technique
has been used here to identify photons from (anti)\Sgm
decays in conjunction with STAR's excellent PID capabilities
for finding the associated (anti)\Lam daughters.
We report here on progress toward measuring the
(anti)\Sgm yields in various nuclear collisions at
RHIC.

\end{abstract}

\submitto{\JPG}

\section{Introduction}

Reasons to understand the production of \GSgm states are various, but generally
stem from their decay contributions to other baryon states. A primary example
of this is the contribution of the \Sgm to the \Lam yields and spectra via its
electromagnetic decay $\Sigma^0 \rightarrow \Lambda \gamma$
(with a branching ratio close to 100\%).  
Because the decay is electromagnetic, the short decay lengths
make this measurement possible only by identification of
the partner \Gm, a task few detectors are designed to handle well.

Awareness of such contributions may be necessary to make accurate
comparisons with particle production models of nuclear collisions. A notable instance involves
one of the more prominent expectations of quark gluon plasma:
strangeness enhancement in the antihyperon channel measured via
$\overline{\Lambda}/\overline{p}$~\cite{enhance}.
Aside from correcting for feeddown from \ALam,
$\overline{\Xi}$, and $\overline{\Omega}$
decays, and even assuming the measured \ALam yield is inclusive
of the \ASgm yield, the observed $\overline{p}$ yield is
still influenced by $\overline{\Sigma}^-$ decays.
Without knowledge of the \GSgm states' yields, unproven
model assumptions must be used to correct for their contributions, leading
to possibly premature conclusions.

Another example of where ignorance of \GSgm production may affect
physics interpretation is in the spectra of the \Lam. Such spectra are often
used in diagnoses of radial flow in heavy ion collisions through
$<$$p_T$$>$ or inverse slope determinations, but certainly include the
products of the \Sgm decay (even though
the electromagnetic and weak decay lifetimes of the \Sgm and $\Sigma^{\pm}$
respectively exclude their decay during
any reasonable lifetime expectations of the fireball).
It is unknown whether final state interactions of the  \GSgm
states are equivalent to that of the \Lam (one expects
at least a small change from their mass difference in the view of ideal flow)
as interaction cross sections with other species have been scarcely
measured (total cross sections of $\Sigma^-$ and \Lam
with protons have been measured with a very
small region of overlap at $\sqrt{s} \approx 15$ GeV, where they
appear to agree with a value of approximately 33 mb~\cite{PDG1}).
Additionally, there exist data from only one experiment on how the
production of \Sgm and \Lam differ over phase space in nuclear collisions
($p$+$Be$ at $p_{lab} = 28.5$ GeV/$c$), from which it appears that
the ratio $\Lambda_{\Sigma^0 \rightarrow \Lambda\gamma}/\Lambda_{inclusive}$
is relatively constant at approximately 1/4.~\cite{pBe}. However,
with possibly different feeddown contributions
from resonance states in larger colliding systems,
that may not be true in heavy ion collisions.

The latter point also has bearing on whether trends observed in the data
are coincidental. Of note, a linearity between \Lam yields and $h^-$
was seen in 130 GeV $Au$+$Au$ data at RHIC~\cite{Lam1}. If the
factors that make up this \Lam  measurement do not scale,
we may be fooled.

\subsection{\Ratio}

From the above arguments, it is evident that a measurement of the
relative yields of the \Sgm and \Lam would prove useful at RHIC.
It has been stated that isospin dictates the ratio of the production
cross sections $\sigma(\Sgmm)/\sigma(\Lambda)$ should
be 1/3~\cite{cosy11}. Reasons for this are not obvious,
so we will examine other production arguments here.
To be clear, we will henceforth focus on such yields after strong
(resonance) decays, but before electroweak decays.

\subsubsection{Thermal Models}

For thermal models, one needs to establish the parameters of the model
to use for determining a ratio. A good example is the set from a fit to
central 200 GeV $Au$+$Au$ data from STAR~\cite{Olga}. Using these parameters
as input to the THERMUS thermal model~\cite{THERMUS}, one obtains for
the primordial ratio a value of 0.67, and for the ratio after resonance decay
contributions a value of 0.36.  The dependence of either of these values on the
parameters of the thermal model is very weak, but it is clear that the
resonances play a strong role. This is noteworthy because there are
indications that these resonances are under-populated relative to the thermal
model in the $Au$+$Au$ data, but are reasonable in $p$+$p$~\cite{Olga,Markert}. So reality
may be somewhere between the two values
if the thermal predictions are correct.

\subsubsection{Quark Coalescence Models}

Simple counting rules in quark coalescence models, expected to be relevant only under
conditions of very dense matter with partonic degrees of freedom~\cite{qcoal,ALCOR},
lead to yield ratios of 1/1 for primordial production, but feeddown from
resonances into both species reduces this ratio to 1/5 if the
resonances are fully-populated~\cite{Levai}. Again, reality may be somewhere
in the middle, but the window is larger than for the thermal model.
If central $Au$+$Au$ collisions produce a \Ratio value near 0.2 or 1.0, 
this may serve as an indicator for quark coalescence.

\subsubsection{Event Generators}

As one example of an event generator,
HIJING/$B\overline{B}$ has no final state rescattering to alter yields
of resonances~\cite{HIJING}. It predicts a value of 0.37 for the ratio in 200 GeV
$d$+$Au$ collisions, which agrees with the thermal model prediction for
a fully equilibrated system at RHIC, including resonances.

\subsubsection{Past Results}

\begin{figure}
\begin{center}
\epsfxsize=5.5in
\epsfbox{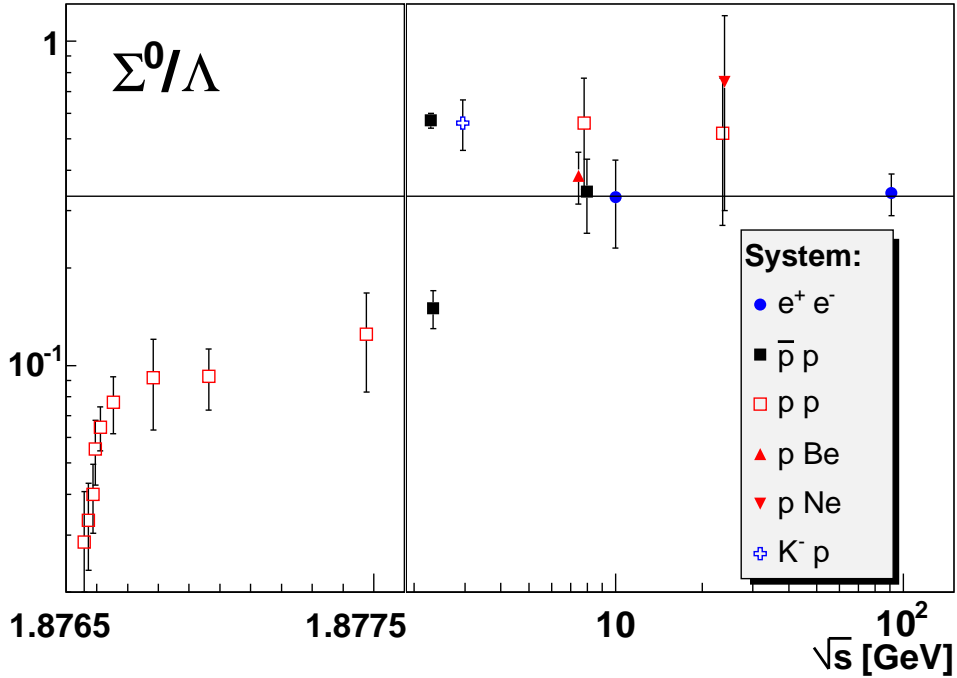}
\vspace{-0.8cm}
\end{center}
\caption{\label{fi:PastRes}\Ratio past results versus collision $\sqrt{s}$
($\sqrt{s_{NN}}$ for $p$+$A$)~\cite{pBe,past,pNe,Z0}.}
\end{figure}

The energy threshold for producing a \Sgm (via associated production,
$p$+$p \rightarrow p K^+ \Sgmm$) is slightly higher than that
for creating a \Lam ($p$+$p \rightarrow p K^+ \Lambda$),
so the initial behavior at low energy is dominated by that of
the threshold for \Sgm production, as shown in Fig.~\ref{fi:PastRes}.
Further details of
production (and of the rise in the \Ratio ratio) at these low energies appear to be well-understood~\cite{lowtheory}.
This is followed perhaps by fluctuations in the ratio when the thresholds
for $\Lambda\overline{\Lambda}$ and $\Sgmm\ASgmm$ pair production are reached,
but there is little data in this range to clarify this.

The expectation is that the data then approach the
"ideal" ratio of 1/3 from isospin as the available energy for production
rises well above thresholds. For the most part, this seems to
be approximately true, though only one data point exists from a colliding system
larger than $p$+$Be$, from $p$+$Ne$ at $\sqrt{s_{NN}} = 24$ GeV~\cite{pNe}.
This point has a
rather high value of $0.75 \pm 0.45$, but is consistent with 1/3 within its errors.
This is insufficient to conclude whether larger colliding systems behave
differently. At higher energies, there exist only the measurements
of $Z^0$ decay products~\cite{Z0}, and nothing from hadronic interactions.

\section{Experiment}

\subsection{Technique}

In order to observe the charged \GSgm states in an experiment, it is
important to have either good \Piz or neutron detection efficiency. The
STAR detector has not had strong capabilities in these areas, though
a measurement has been made of the \Piz spectrum using
the STAR TPC detector in 130 GeV $Au$+$Au$ collisions~\cite{pi0}.
It may also be possible to use the now-operational
STAR electromagnetic calorimeters to identify \Piz candidates
for (anti)$\Sigma^+$ decays~\cite{emc}.

The STAR TPC \Piz measurement took advantage of photon conversions
($\gamma \rightarrow e^+e^-$) in detector material to observe the
4-body final state $\pi^0 \rightarrow \gamma \gamma \rightarrow e^+e^-e^+e^-$.
While the combinatorics of the technique leave the background under
the \Piz mass peak too large for identifying individual pions, the reconstructed
photons do have enough signal-to-noise to use in combination with
other potential decay products (as demonstrated by the \Piz measurement).
This, along with strong \Lam identification capabilities,
is what makes \Sgm decay observation possible in STAR.

\begin{figure}
\begin{center}
\epsfxsize=4.0in
\epsfbox{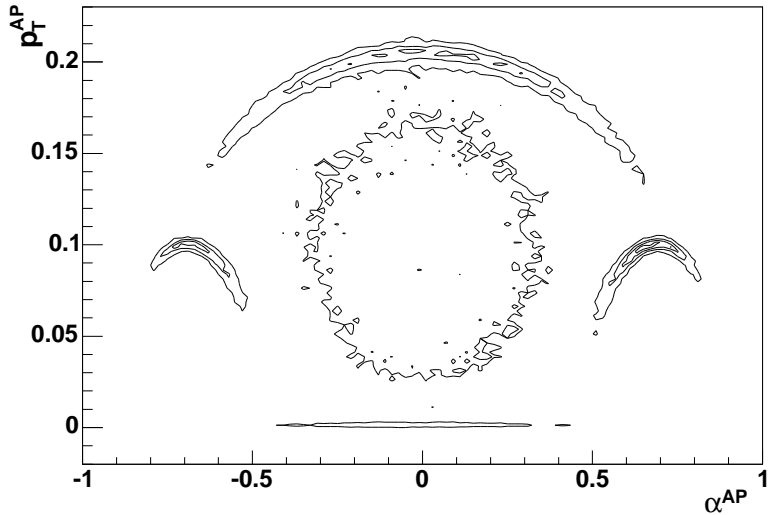}
\vspace{-0.7cm}
\end{center}
\caption{\label{fi:armen}Contour plot of
Armenteros-Podolanski decay variables \ptarm\ versus $\alpha^{AP}$
from $V0$ decays in the STAR detector. Notable features include
combinatoric background in the center, a $K^0_{short}$ arc across
the top, (anti)\Lam arcs on the sides, and a \Gm\ conversion
band along the bottom at low \ptarm.}
\end{figure}

\begin{figure}
\begin{center}
\epsfxsize=6.3in
\epsfbox{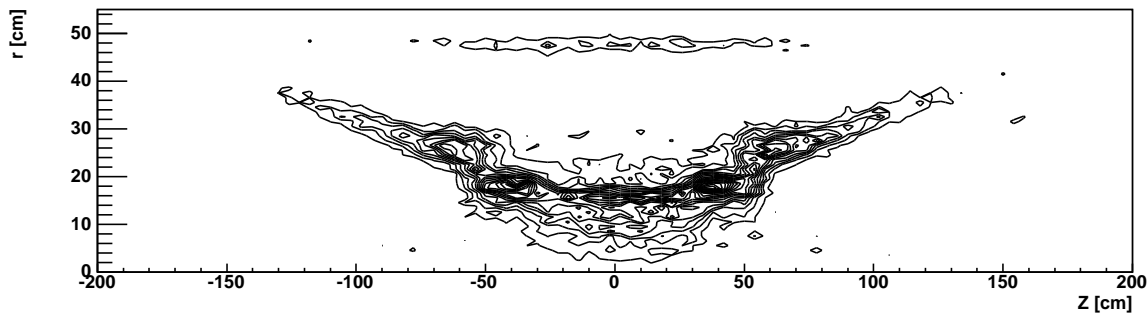}
\vspace{-1.3cm}
\end{center}
\caption{\label{fi:Conversions}Contour plot of
conversion locations of \Gm\ candidates in radius versus z (the beam axis),
revealing the mechanical structure of the STAR detector.}
\end{figure}

In the current \Sgm analysis, a different \Gm-finder was used from that
of the \Piz analysis. It turns out that a photon conversion
has all the characteristics of the $V0$ decays used in weak decay
analyses in STAR: two oppositely charged particles originating from a displaced
vertex (several centimeters or more) with a reconstructed parent originating
from the primary collision vertex.  However, two aspects uniquely identify
these $V0$s: their daughters have $d$E$/dx$  characteristics of electrons,
and they occupy a unique region of the Armenteros-Podolanski decay variable
\ptarm\ (shown in Fig.~\ref{fi:armen}).
Because the electron $d$E$/dx$ band crosses other
$d$E$/dx$ bands in the range of interest (low momentum),
a simple cut on low \ptarm\ is the most stringent. 
The conversions are most likely to occur in detector material, so
the locations of the \Gm\ candidates' decay vertices clearly reveal
the structure of the STAR apparatus~\cite{STARdet}, as can be seen in Fig.~\ref{fi:Conversions}. 

\subsection{Data}

The combinatoric background in 200 GeV $Au$+$Au$ collisions
has proven to be too difficult to overcome
using this reconstruction technique, and no clear signal has been seen. While this
is not an issue in $p$+$p$ data in STAR, the accumulated statistics are too few
to perform an analysis.

\begin{figure}
\begin{center}
\epsfxsize=4.1in
\epsfbox{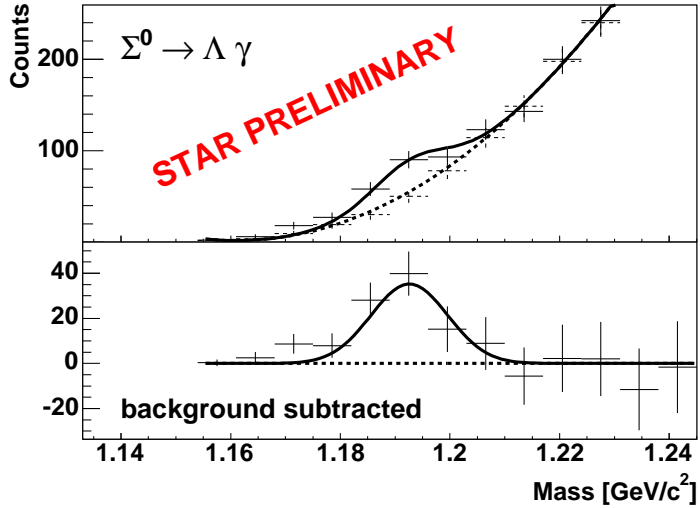}
\vspace{-0.7cm}
\end{center}
\caption{\label{fi:data}Reconstucted \Sgm
decay invariant mass distribution (solid) and
rotated background approximation (dashed) with fits,
before and after background subtraction,
from mid-rapidity minimum bias 200 GeV $d$+$Au$ data in STAR.}
\end{figure}

The STAR data collected from minimum bias 200 GeV $d$+$Au$ collisions, however, benefit from
both low combinatoric backgrounds, and significant statistics. From 14.7 million
minimum bias events, one can see clear evidence of the \Sgm signal
as shown in Fig.~\ref{fi:data} for rapidities between $\pm 0.75$
over all $p_T$.

\begin{figure}
\begin{center}
\epsfxsize=4.5in
\epsfbox{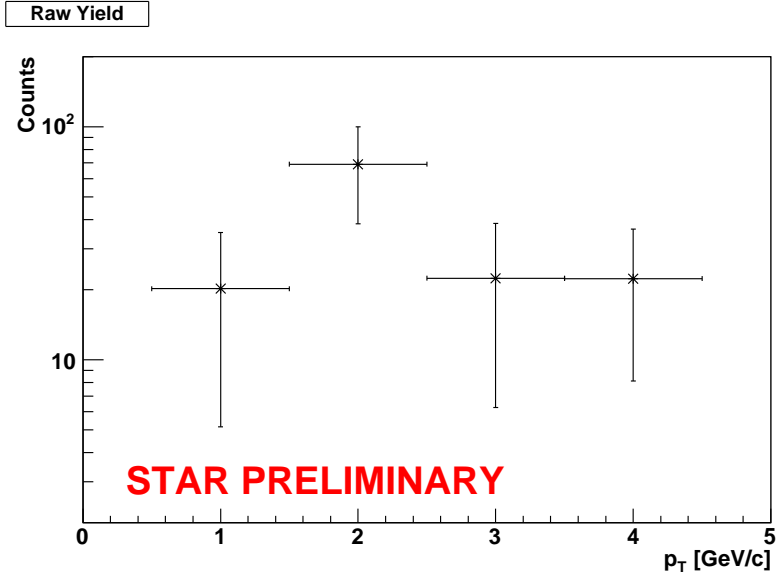}
\vspace{-0.8cm}
\end{center}
\caption{\label{fi:counts}Reconstucted counts of $\Sgmm$+$\ASgmm$
decays from mid-rapidity minbias 200 GeV $d$+$Au$ data in STAR.}
\end{figure}

To measure a yield, it is necessary to understand the background under
the invariant mass peaks. For this purpose, an approximation of the
background shape is determined by rotations of the \Lam
and \Gm\ candidates with respect to each other in azimuth
(about the beam axis). This technique ensures that any possible
event classes (such as number of binary collisions) are appropriately
represented. Within the errors of the data points, the approximation
appears roughly consistent when scaled to match the data away from the
invariant mass region of interest. This provides confidence that one
can fit the shape of the background from the approximation, and
include some representation of the signal on top of it. Varying the
fit functions and ranges gives a systematic error, but the statistical
errors of the measurements dominate. The resulting reconstructed counts
from four $p_T$ bins of $\Sgmm$+$\ASgmm$
are shown in Fig.~\ref{fi:counts}. Using the uncorrected counts over
all $p_T$ we find $\ASgmm/\Sgmm = 0.6 \pm 0.3$.

\subsection{Corrections}

Efficiency corrections for this \Sgm analysis are difficult to certify. Statistics
under the invariant mass peak are so few that comparisons cannot be made
between real and simulated data for distributions used for cuts. Rather than
tackle this issue, the analysis is presently being modified to use the same
\Gm-finder which was used in the \Piz analysis, which has shown
increased efficiencies for identifying \Gm\ candidates from \Piz decays over
simply selecting $V0$ candidates with low  \ptarm.
Efficiencies and yields from the modified analysis are expected
in the future.

The experimentally measurable \Sgm yield suffers from only minor feeddown
corrections. From weak decays, contributions include $\Xi^0$ decays with
a branching ratio of $(3.33 \pm 0.10) \times 10^{-3}$, and $\Xi^-$ decays
with branching ratios below $10^{-3}$. Even if $\Xi$ yields were similar to
that of the \Lam, the feeddown to \Sgm would be small.
Geometric cuts on the \Sgm originating from the collision vertex
also cut mildly into the efficiency for reconstruction of these secondary \Sgm
candidates.

Strong decays into the \Sgm include a $12 \pm 2\%$ branching ratio from $\Sigma$(1385)
decays and undetermined fractions of other, heavier excited \GSgm states. Excited \Lam
states can also play a role with 100\% and 40\% of $\Lambda$(1405) and $\Lambda$(1520)
decaying into $\Sigma \pi$, respectively, and uncertain fractions of higher mass states.
But the expectedly lower yields of all such excited states implies only small contributions
from the lightest of these. Measurements of the $\Sigma$(1385) yield
in high energy nuclear collisions by STAR~\cite{sevil} will likely aid our understanding of
possible resonance feeddown corrections.

\section{Summary}

We have observed a signal for (anti)\Sgm in STAR for
minimum bias 200 GeV $d$+$Au$ collisions at midrapidity using a 4-body final state
reconstruction. While the signal is not
strong, there is hope for improvement via a different \Gm-finder
which has shown higher efficiencies in other STAR analyses, such that
a yield can be measured. Finding yields in other colliding
systems may also be possible with further data acquired by STAR:
improved statistics of 200 GeV $Au$+$Au$ data from 2004 may overcome
the large combinatorial background in that data; 62 GeV $Au$+$Au$ data
taken in 2004 is expected to have reduced combinatorial background,
as should $Cu$+$Cu$ data to be taken in 2005; and future $p$+$p$ runs should provide
sufficient statistics to observe the (anti)\Sgm signals.


\section*{References}
\begin{thebibliography}{11}
\bibitem{sevil}Salur S (STAR Collaboration) 2004 {\it these proceedings}
\bibitem{pi0}Adams J \etal (STAR Collaboration) 2004 {\it Phys. Rev.} C {\bf 70} 044902
\bibitem{enhance}Koch P, Rafelski J, and Greiner W 1983 {\it Phys. Lett.} B {\bf 123} 151
\bibitem{PDG1}Eidelman S \etal 2004 {\it Phys. Lett.} B {\bf 592} 1
\bibitem{pBe}Sullivan M W \etal 1987 {\it Phys. Rev.} D {\bf 36} 674
\bibitem{Lam1}Adler C \etal 2002 {\it Phys. Rev. Lett} {\bf 89} 092301
\bibitem{cosy11}Kowina P \etal (COSY-11 Collaboration) 2004 {\it Preprint} nucl-ex/0401020
\bibitem{Olga}Barannikova O (STAR Collaboration) 2004 {\it these proceedings}
\bibitem{THERMUS}Wheaton S and Cleymans J 2004 {\it Preprint} hep-ph/0407174
\bibitem{Markert}Markert C 2004 {\it these proceedings}
\bibitem{qcoal}Bir\'{o} T S and Zim\'{a}nyi J 1983 {\it Nucl. Phys.} A {\bf 395} 525
\bibitem{ALCOR}Bir\'{o} T S, L\'{e}vai P, and Zim\'{a}nyi J 1995 {\it Phys. Lett.} B {\bf 347} 6
\bibitem{Levai}L\'{e}vai 2004 P {\it private communications}
\bibitem{HIJING}Vance S E, Gyulassy M, Wang X N 1998 {\it Phys. Lett} B {\bf 443} 45
\bibitem{past}Albrecht H (ARGUS Collaboration) 1987 {\it Phys. Lett.} B {\bf 183} 419;
Bogolyubsky M Y \etal 1986 {\it Yad. Fiz.} {\bf 43} 1199;
Bogolyubsky M Y \etal 1989 {\it Yad. Fiz.} {\bf 50} 683;
Baubillier M \etal 1978 {\it Nucl. Phys.} B {\bf 148} 18;
Eisenstein R A (PS185 Collaboration) 1994 {\it Phys. Atom. Nucl.} {\bf 57} 1680;
Kowina P \etal (COSY-11 Collaboration) 2004 {\it Preprint} nucl-ex/0401020;
Sewerin S \etal (COSY-11 Collaboration) 1999 {\it Phys. Rev. Lett.} {\bf 83} 682
\bibitem{lowtheory}Sibirtsev A \etal 2000 {\it Preprint} nucl-th/0004022;
Shyam R, Penner G, and Mosel U 2001 {\it Phys. Rev.} C {\bf 63} 022202;
Gasparian A \etal 2001 {\it Nucl. Phys.} A {\bf 684} 397.
\bibitem{pNe}Yuldashev B S \etal (FNAL-343 Collaboration) 1991 {\it Phys. Rev.} D {\bf 43} 2792
\bibitem{Z0}Acciarri M \etal (L3 Collaboration) 1994 {\it Phys. Lett.} B {\bf 328} 223;
Acciarri M \etal (L3 Collaboration) 2000 {\it Phys. Lett} B {\bf 479} 79
\bibitem{emc}Mischke A (STAR Collaboration) 2004 {\it Preprint} nucl-ex/0410032
\bibitem{STARdet}Ackermann K H \etal (STAR Collaboration) 2003 {\it Nucl. Inst. and Meth.} A {\bf 499} 624
\end{thebibliography}
\end{document}